\documentstyle[aps,epsf,rotate,preprint]{revtex}
\begin{document}
 
\draft

\title{Universal and non-universal properties of cross-correlations \\
in financial time series}

\author{Vasiliki Plerou$^{1,2}$, Parameswaran Gopikrishnan$^{1}$, 
Bernd Rosenow$^{3}$, \\ Lu\'{\i}s~A.~Nunes Amaral$^{1}$, 
and H.~Eugene Stanley$^1$}

\address{
$^1$Center for Polymer Studies and Department of Physics,
        Boston University, Boston, MA 02215 \\
$^2$Department of Physics,
        Boston College, Chestnut Hill, MA 02167 \\
$^3$Institut f\"ur Theoretische Physik, Universit\"at zu K\"oln, 
        D--50937 K\"oln, Germany \\
}

\date{Last modified: Feb 14, 1999; Printed: \today}

\maketitle

\begin{abstract}
  
  We use methods of random matrix theory to analyze the cross-correlation
  matrix {\bf \sf C} of price changes of the largest 1000 US stocks for the
  2-year period 1994-95.  We find that the statistics of most of the
  eigenvalues in the spectrum of {\bf \sf C} agree with the predictions of
  random matrix theory, but there are deviations for a few of the largest
  eigenvalues.  We find that {\bf \sf C} has the universal properties of the
  Gaussian orthogonal ensemble of random matrices. Furthermore, we analyze
  the eigenvectors of {\bf \sf C} through their inverse participation ratio
  and find eigenvectors with large inverse participation ratios at both edges
  of the eigenvalue spectrum---a situation reminiscent of results in
  localization theory.

\end{abstract}




There has been much recent work applying physics concepts and methods to
the study of financial time
series~\cite{Mandelbrot,econo,Lux,Bouchaud,Zhang,Takayasu,dietrich,BU,
Sornette,bou1,Mantegna,Solomon,Ghasghaie,Ausloos}.  In particular, the
study of correlations between price changes of different stocks is both
of scientific interest and of practical relevance in quantifying the
risk of a given stock portfolio~\cite{Mandelbrot,econo}. Consider, for
example, the equal-time correlation of stock price changes for a given
pair of companies.  Since the market conditions may not be stationary,
and the historical records are finite, it is not clear if a measured
correlation of price changes of two stocks is just due to ``noise'' or
genuinely arises from the interactions among the two
companies. Moreover, unlike most physical systems, there is no
``algorithm'' to calculate the ``interaction strength'' between two
companies (as there is for, say, two spins in a magnet). The problem is
that although every pair of companies should interact either directly or
indirectly, the precise nature of interaction is unknown.

In some ways, the problem of interpreting the correlations between
individual stock-price changes is reminiscent of the difficulties
experienced by physicists in the fifties, in interpreting the spectra of
complex nuclei.  Large amounts of spectroscopic data on the energy
levels were becoming available but were too complex to be explained by
model calculations because the exact nature of the interactions were
unknown.  Random matrix theory (RMT) was developed in this context, to
deal with the statistics of energy levels of complex quantum
systems\cite{wigner,dyson}. With the minimal assumption of a random
Hamiltonian, given by a real symmetric matrix with independent random
elements, a series of remarkable predictions were made and successfully
tested on the spectra of complex nuclei~\cite{wigner}. RMT predictions
represent an average over all possible
interactions~\cite{dyson}. Deviations from the {\it universal\/}
predictions of RMT identify system-specific, non-random properties of
the system under consideration, providing clues about the nature of the
underlying interactions~\cite{review,eigcorr}.


In this letter, we apply RMT methods to study the
cross-correlations~\cite{bou1} of stock price changes. First, we
demonstrate the validity of the universal predictions of RMT for the
eigenvalue statistics of the cross-correlation matrix.  Second, we
calculate the deviations of the empirical data from the RMT predictions,
obtaining information that enables us to identify cross-correlations
between stocks not explainable purely by randomness.

We analyze a data base~\cite{TAQ} containing the price $S_i(t)$ of stock
$i$ at time $t$, where $i = 1, \dots , 1000$ denotes the largest 1000
publicly-traded companies and the time $t$ runs over the 2-year period
1994-95.  From this time series, we calculate the price change
$G_i(t,\Delta t)$, defined as
%
\begin{equation}
G_i (t,\Delta t) \equiv \ln S_i(t+\Delta t) - \ln S_i (t) \,,
\label{eq.1}
\end{equation}
%
where $\Delta t=30$~min is the sampling time scale.  The simplest
measure of correlations between different stocks is the equal-time
cross-correlation matrix {\bf \sf C} which has elements
%
\begin{equation}
C_{ij} \equiv { \langle G_i G_j \rangle - \langle G_i \rangle 
\langle G_j \rangle \over \sigma_i \sigma_j } \,,
\label{eq.2}
\end{equation}
%
where $\sigma_i \equiv \sqrt{\langle G_i^2 \rangle - \langle G_i
\rangle^2} $ is the standard deviation of the price changes of company
$i$, and $\langle\cdots\rangle$ denotes a time average over the period
studied\cite{TAQ}.

\begin{figure}
\narrowtext
\centerline{
\epsfysize=0.8\columnwidth{\rotate[r]{\epsfbox{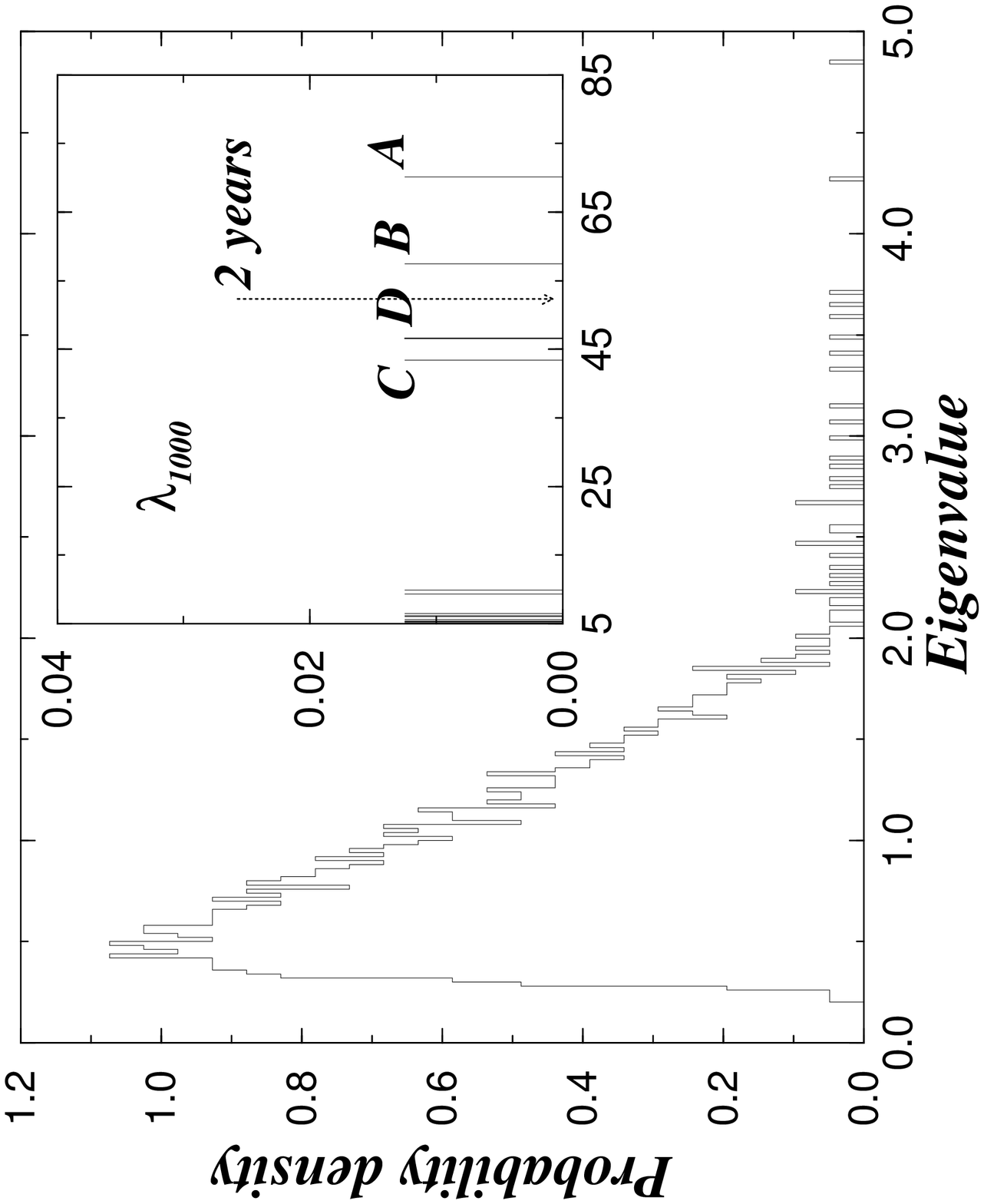}}}}
\vspace*{0.5cm}
\caption{ The probability density of the eigenvalues of the normalized
  cross-correlation matrix {\bf \sf C} for the 1000 largest stocks in the TAQ
  database for the 2-year period 1994-95 \protect\cite{TAQ}. Recent
  analytical results~\protect\cite{sengupta} for cross-correlation matrices
  generated from uncorrelated time series predict a finite range of
  eigenvalues depending on the ratio $R$ of the length of the time series to
  the dimension of the matrix~\protect\cite{bou1}.  In our case $R=6.448$
  corresponding to eigenvalues distributed in the interval $0.37 \leq
  \lambda_k \leq 1.94$~\protect\cite{sengupta}. However, the largest
  eigenvalue for the 2-year period (inset) is approximately 30 times larger
  than the maximum eigenvalue predicted for uncorrelated time series. The
  inset also shows the largest eigenvalue for the cross-correlation matrix
  for 4 half-year periods---denoted A, B, C, D. The arrow in the inset
  corresponds to the largest eigenvalue for the entire 2-year period,
  $\lambda_{1000} \approx 50$. The distribution of eigenvector components for
  the large eigenvalues, well outside the bulk show significant deviations
  from the Gaussian prediction of RMT, which suggests ``collective'' behavior
  or correlations~\protect\cite{eigcorr} between different companies. The
  largest eigenvalue would then correspond to the correlations within the
  entire market~\protect\cite{bou1}.}
\label{f.density}
\end{figure}

We analyze the statistical properties of {\bf \sf C} by applying RMT
techniques. First, we diagonalize {\bf \sf C} and obtain its eigenvalues
$\lambda_k$ ---with $k=1,\cdots,1000$---which we rank-order from the
smallest to the largest. Next, we calculate the eigenvalue
distribution~\cite{bou1} and compare it with recent analytical results
for a cross-correlation matrix generated from finite uncorrelated time
series~\cite{sengupta}. Figure~\ref{f.density} shows the eigenvalue
distribution of {\bf \sf C}, which deviates from the predictions of
Ref.~\cite{sengupta}, for large eigenvalues $\lambda_k \ge 1.94$ (see
caption of Fig.~\ref{f.density}). This result is in agreement with the
results of Ref.~\cite{bou1} for the eigenvalue distribution of {\bf \sf
C} on a daily time scale.

\begin{figure}
\narrowtext
\centerline{
\epsfysize=0.8\columnwidth{\rotate[r]{\epsfbox{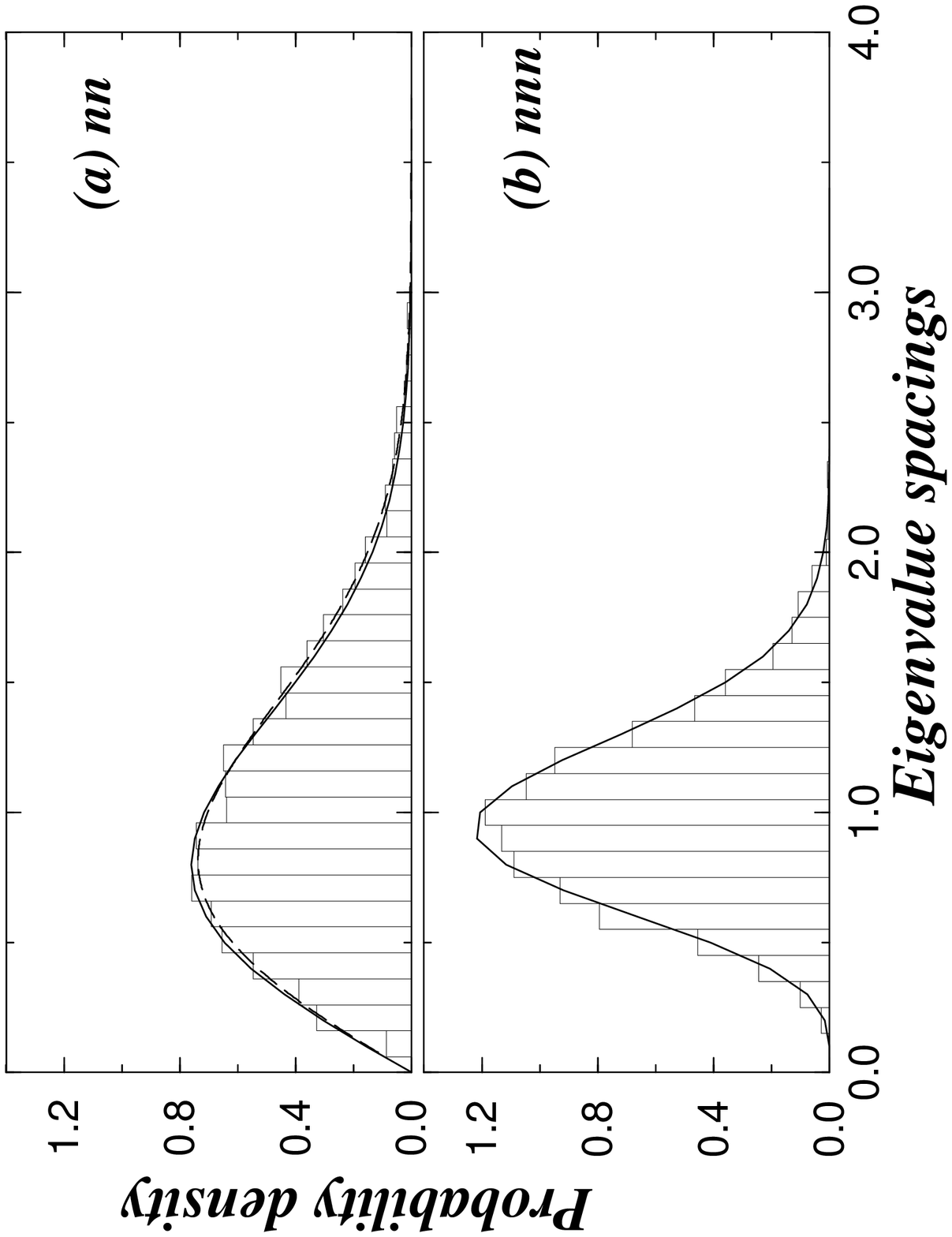}}}}
\vspace*{1.0cm}
\centerline{
\epsfysize=0.8\columnwidth{\rotate[r]{\epsfbox{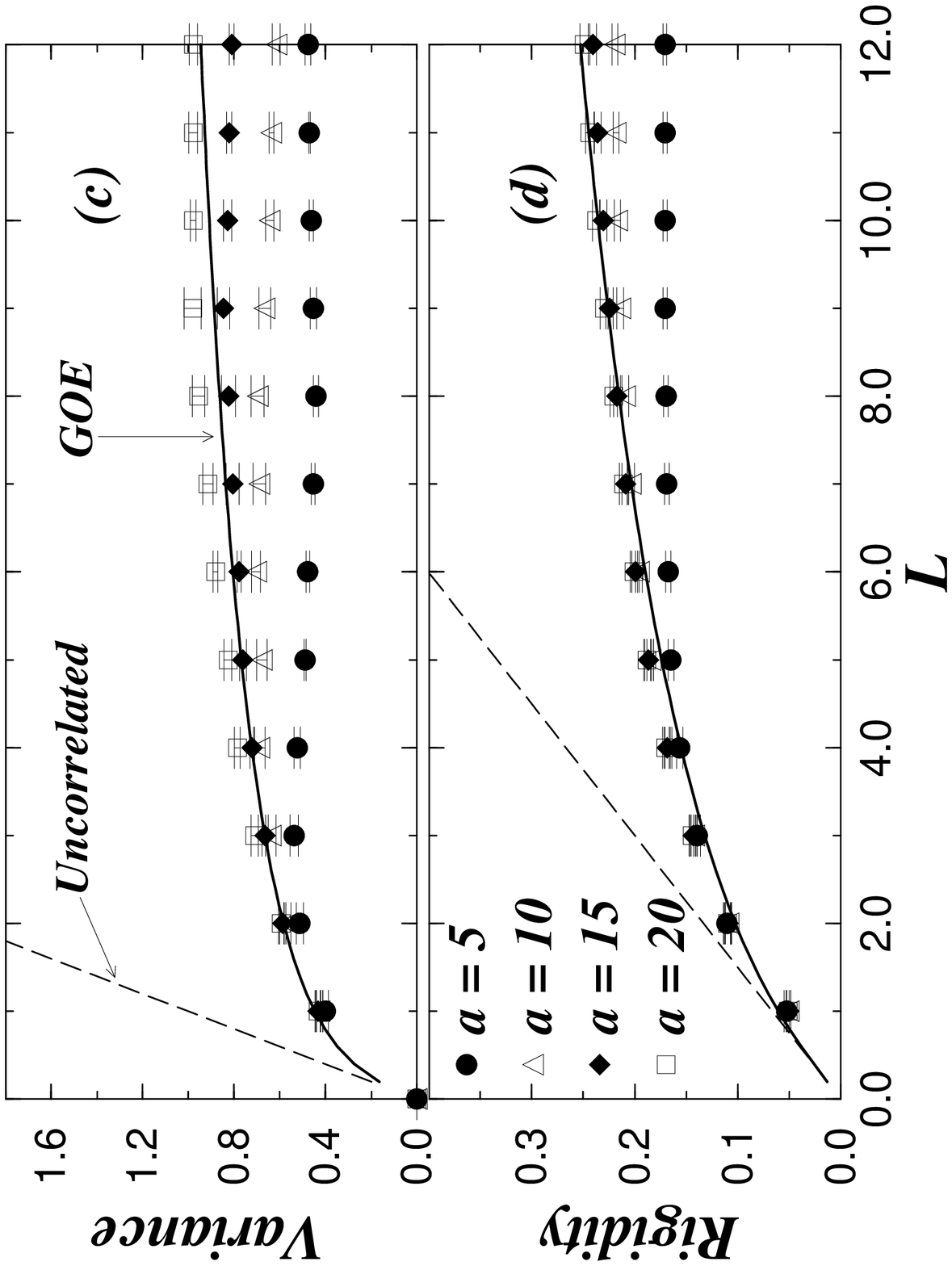}}}}
\vspace*{1.0cm}
\caption{ Comparison of the RMT predictions for the spacing
distributions with results for empirical cross-correlation matrix . (a)
Nearest-neighbor ({\it nn}) spacing distribution of the eigenvalues of
{\bf \sf C} after unfolding. We use the Gaussian broadening
procedure~\protect\cite{unfold}. The eigenvalue distribution can be
considered as a sum of delta functions about each eigenvalue,
$\lambda_k$, each of which is then ``broadened'' by choosing a Gaussian
distribution with standard deviation $(\lambda_{k+a} - \lambda_{k-a})/2
$, where $2a$ is the size of the window used for
broadening~\protect\cite{unfold}. Here, $a=15$, the optimum value
obtained from Fig.~\protect\ref{f.nearest}(d). The solid line is the GOE
prediction, Eq.~(\protect\ref{eq.3}), and the dashed line is a fit to
the one parameter Brody distribution $ p(s) \equiv B\,(1+\beta)\,
s^{\beta}\, \exp(-B s^{\beta+1})$, with $B\equiv [\Gamma ({\beta+2 \over
\beta+1})]^{1+\beta}$.  The fit yields $\beta = 0.99 \pm 0.02$, in good
agreement with the GOE prediction $\beta=1$. A Kolmogorov-Smirnov test
suggests that the GOE is $10^{5}$ times more likely to be the correct
description than the Gaussian unitary ensemble, and $10^{20}$ times more
likely than the GSE.  Furthermore, at the 80\% confidence level, the
Kolmogorov-Smirnov test cannot reject the hypothesis that the GOE is the
correct description.  (b) Next-nearest-neighbor ({\it nnn}) spacing
distribution of {\bf \sf C}. RMT predicts that, for the GOE, the
distribution of next-nearest-neighbor spacing should follow the same
distribution as the nearest-neighbor spacing for the GSE.  This
prediction is confirmed for the empirical data both visually and by a
Kolmogorov-Smirnov test that at the 40\% confidence level cannot reject
the hypothesis that the GSE is the correct distribution.  (c) Number
variance and (d) spectral rigidity of {\bf \sf C } for different values
of the unfolding parameter $a$, as compared to the exact expression for
the GOE (solid line) and the uncorrelated case (dashed line) . As $a$
increases, both the number variance and the spectral rigidity approach
the theoretical curve for the GOE while the spacing distribution remains
essentially unchanged. We choose $a=15$ as the optimal-value. }
\label{f.nearest}
\end{figure}

To test for universal properties, we first calculate the
distribution of the nearest-neighbor spacings $s\equiv \lambda_{k+1}
-\lambda_k$. The nearest-neighbor spacing is computed after transforming
the eigenvalues in such a way that their distribution becomes
uniform---a procedure known as unfolding\cite{review,eigcorr,unfold}.
Figure~\ref{f.nearest}(a) shows the distribution of nearest-neighbor
spacings for the empirical data, and compares it with the RMT
predictions for real symmetric random matrices.  This class of matrices
shares universal properties with the ensemble of matrices whose elements
are distributed according to a Gaussian probability measure---the
Gaussian orthogonal ensemble (GOE). We find good agreement between the
empirical data and the GOE prediction,
%
\begin{equation}
P_{\rm GOE}(s)= {\pi s \over 2}\, \exp\left(- {\pi \over 4}\, 
s^2 \right)\,.
\label{eq.3}
\end{equation}
%

A second independent test of the GOE is the distribution of {\it
next}-nearest-neighbor spacings between the rank-ordered
eigenvalues~\cite{review}. This distribution is expected to be identical
to the distribution of nearest-neighbor spacings of the Gaussian
symplectic ensemble (GSE) as verified by the empirical data
[Fig.~\ref{f.nearest}(b)].

The distribution of eigenvalue spacings reflects correlations only of
consecutive eigenvalues but does not contain information about
correlations of longer range. To probe any ``long-range'' correlations,
we first calculate the number variance $\Sigma^2$ which is defined as
the variance of the number of unfolded eigenvalues in intervals of
length $L$ around each of the
eigenvalues\cite{review,eigcorr,unfold,numvardef}. If the eigenvalues
are uncorrelated, $\Sigma^2 \sim L$. For the opposite case of a
``rigid'' eigenvalue spectrum, $\Sigma^2$ is a constant. For the GOE
case, we find the ``intermediate'' behavior $\Sigma^2 \sim \ln L$, as
predicted by RMT [Fig.~\ref{f.nearest}(c)].

A second way to measure ``long-range'' correlations in the eigenvalues
is through the spectral rigidity $\Delta$, defined to be the least
square deviation of the unfolded cumulative eigenvalue density from a
fit to a straight line in an interval of length
$L$\cite{review,eigcorr,unfold,rigdef}.  For uncorrelated eigenvalues,
$\Delta \sim L$, whereas for the rigid case $\Delta$ is a constant. For
the GOE case we find $\Delta \sim \ln L$ as predicted by RMT
[Fig.~\ref{f.nearest}(d)].

Having demonstrated that the eigenvalue statistics of {\bf \sf C}
satisfies the RMT predictions, we now proceed to analyze the
eigenvectors of {\bf \sf C}.  RMT predicts that the components of the
normalized eigenvectors of a GOE matrix are distributed according to a
Gaussian probability distribution with mean zero and variance one. In
agreement with recent results~\cite{bou1}, we find that eigenvectors
corresponding to {\it most\/} eigenvalues in the ``bulk'' ($\lambda_k
\leq 2$) follow this prediction. On the other hand, eigenvectors with
eigenvalues outside the bulk ($\lambda_k \ge 2$) show marked deviations
from the Gaussian distribution.  In particular, the vector corresponding
to the largest eigenvalue $\lambda_{1000}$ deviates significantly from
the Gaussian distribution predicted by RMT.

The component $\ell$ of a given eigenvector relates to the contribution
of company $\ell$ to that eigenvector.  Hence, the distribution of the
components contains information about the number of companies
contributing to a specific eigenvector.  In order to distinguish between
one eigenvector with approximately equal components and another with a
small number of large components we define the inverse participation
ratio~\cite{review,bandmatrix}
%
\begin{equation}
I_{k} \equiv  \sum_{\ell=1}^{1000} [u_{k\ell}]^4 \ ,
\label{e.ipr}
\end{equation}
%
where $u_{k\ell}$, $\ell=1,\dots,1000$ are the components of
eigenvector $k$. The physical meaning of $I_k$ can be illustrated by
two limiting cases: (i) a vector with identical components
$u_{k\ell}\equiv 1/\sqrt{N}$ has $I_k=1/N$, whereas (ii) a vector with
one component $u_{k1}=1$ and all the others zero has
$I_k=1$. Therefore, $I_k$ is related to the reciprocal of the number
of vector components significantly different from zero.

\begin{figure}
\narrowtext
\centerline{
\epsfysize=0.8\columnwidth{\rotate[r]{\epsfbox{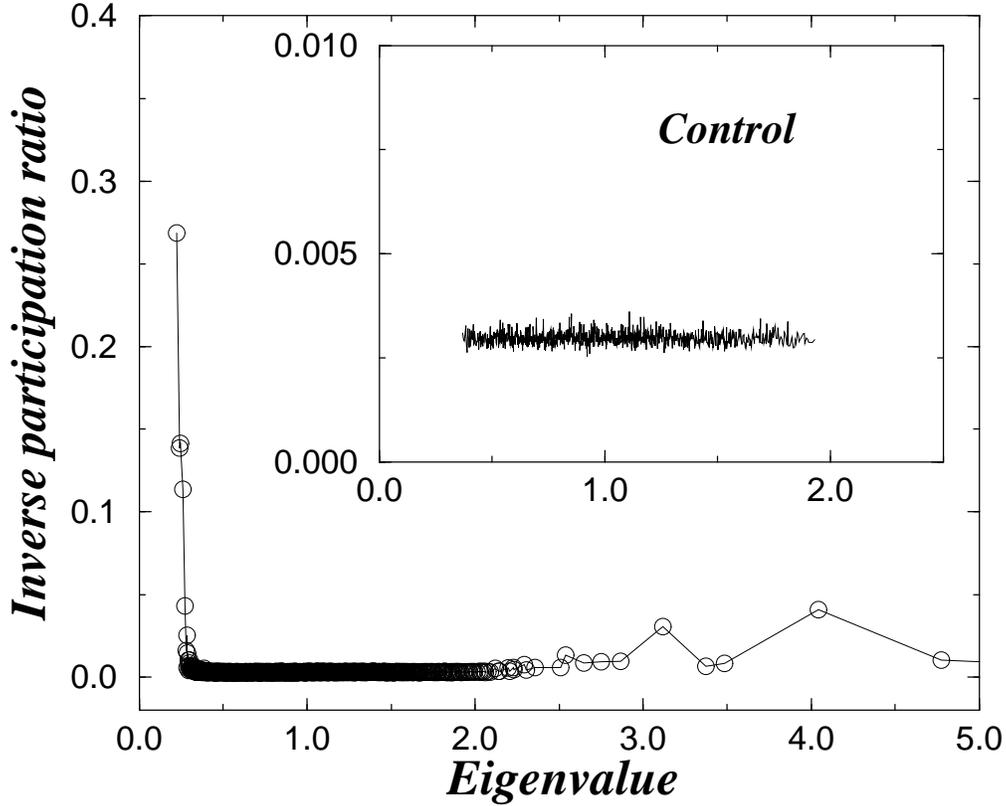}}}
}
\vspace*{0.5cm}
\caption{ Inverse participation ratio $I_k$ for each of the 1000
eigenvectors. As a control, we show in the inset the $I_k$ values for
the eigenvectors of a cross-correlation matrix computed from
uncorrelated independent power-law distributed time
series\protect\cite{BU} of the same length as the data. Empirical data
show marked peaks at both edges of the spectrum, whereas the control
shows only small fluctuations around the average value $ \langle I
\rangle = 3\times 10^{-3}$. The large $I_k$ values for the largest
eigenvalues are to be expected from Fig.~\protect\ref{f.density}, but
the large values of $I_k$ for the small eigenvalues are
surprising. Large $I_k$ values at the edges of the eigenvalue spectrum
is a situation often found in localization theory.}
\label{f.inverse}
\end{figure}

Figure~\ref{f.inverse} shows $I_k$ for eigenvectors of a matrix
generated from uncorrelated time series with a power law distribution of
price changes\cite{BU}.  The average value of $I_k$ is $\langle I
\rangle \approx 3\times10^{-3} \approx 1/N$ indicating that the vectors
are {\it extended\/}~\cite{bandmatrix,extended}---i.e., almost all
companies contribute to them. Fluctuations around this average value are
confined to a narrow range.  On the other hand, the empirical data show
deviations of $I_k$ from $\langle I \rangle$ for a few of the largest
eigenvalues.  These $I_k$ values are approximately 4-5 times larger than
$\langle I \rangle$ which suggests that there are groups of
approximately 50 companies contributing to these eigenvectors.  The
corresponding eigenvalues are well outside the bulk, suggesting that
these companies are correlated~\cite{eigcorr}.

Surprisingly, we also find that there are $I_k$ values as large as
$0.35$ for vectors corresponding to the smallest eigenvalues $\lambda_i
\approx 0.25$~\cite{localization}. These deviations from the average are
two orders of magnitude larger than $\langle I \rangle$, which suggests
that the vectors are {\it localized\/}~\cite{bandmatrix,extended}---i.e.,
only a few companies contribute to them.  The small values of the
corresponding eigenvalues suggests that these companies are uncorrelated
with each other.

The presence of vectors with large $I_k$ also arises in the theory of
Anderson localization\cite{electrons}.  In the context of localization
theory, one frequently finds ``random band matrices''\cite{bandmatrix}
containing extended states with small $I_k$ in the middle of the band,
whereas edge states are localized and have large $I_k$.  Our finding of
localized states for small and large eigenvalues of the
cross-correlation matrix {\bf \sf C} is reminiscent of Anderson
localization and suggests that {\bf \sf C} may be a random band
matrix\cite{rbm}

In summary, we find that the most eigenvalues in the spectrum of the
cross-correlation matrix of stock price changes agree surprisingly well
with the {\it universal} predictions of random matrix theory. In
particular, we find that {\bf \sf C} satisfies the universal properties
of the Gaussian orthogonal ensemble of real symmetric random
matrices. We find through the analysis of the inverse participation
ratio of its eigenvectors that {\bf \sf C} may be a random band matrix,
which may support the idea that a metric can be defined on the space of
companies and that a distance can be defined between pairs of
companies\cite{hierarchy}.  Hypothetically, the presence of localized
states may allow us to draw conclusions about the ``spatial dimension''
of the set of stocks studied here and about the ``range'' of the
correlations between the companies.

We thank M.~Barth\'el\'emy, N.V.~Dohkolyan, X.~Gabaix, U.~Gerland,
S.~Havlin, R.N. Mantegna, Y.~Lee, C.-K.-Peng and D.~Stauffer for helpful
discussions.  LANA thanks FCT/Portugal for financial support.  The
Center for Polymer Studies is supported by NSF.


\begin{references}
\bibitem{numvardef} The number variance is defined as
$\Sigma^2(L)\equiv \langle [N(\lambda+{L \over 2}) -N(\lambda-{L \over
2}) -L]^2 \rangle_{\lambda}$, where $N(\lambda)\equiv \sum_{i} \theta
(\lambda - \lambda_i)$ is the integrated density of eigenvalues and
$\langle\dots\rangle_{\lambda}$ denotes an average over
$\lambda$~\cite{review,unfold}.
\bibitem{rigdef} The spectral rigidity is defined as $\Delta (L)\equiv
{1\over L}\left\langle \min_{A,B}\int_{\lambda-L/2}^{\lambda + L/2}
(N(\lambda_{1}) - A\lambda_{1} -B)^2
d\lambda_{1}\right\rangle_{\lambda}\,,$ where
$\langle\dots\rangle_{\lambda}$ denotes an average over $\lambda$ and
$N(\lambda)\equiv \sum_{i} \theta (\lambda - \lambda_i)$ is the
integrated density of eigenvalues~\cite{review,unfold}.
\bibitem{localization} The large values of $I_k$ for small eigenvalues
indicate that the distribution of eigenvector components for the
eigenvalues at the lower edge of the spectrum deviate from Gaussian
prediction.
\bibitem{hierarchy} A related idea for a hierarchical structure of the
financial cross-correlation matrix was recently put forward in
Ref.\protect\cite{Mantegna}.
\end{references}
\end{document}